\documentclass[prb,twocolumn]{revtex4}
\ifx\pdfoutput\undefined
 \usepackage[dvips]{graphicx}
 \else
 \usepackage[pdftex]{graphicx}
   \fi
   \usepackage{epstopdf}

\usepackage{epsfig}
\usepackage{graphicx}
\usepackage{bm}
\usepackage{amsfonts}

\usepackage[matrix,frame,arrow]{xy}
\usepackage{amsmath}

\newcommand{\Eq}[1]{Eq.~(\ref{#1})}

\newcommand{\Sec}[1]{Sec.~\ref{#1}}
\newcommand{\Fig}[1]{Fig.~\ref{#1}}

\newcommand{\be}{\begin{equation}}
\newcommand{\ee}{\end{equation}}
\newcommand{\bea}{\begin{eqnarray}}
\newcommand{\eea}{\end{eqnarray}}

\newcommand{\bra}[1]{\left\langle \, #1 \right|}
\newcommand{\ket}[1]{\left| #1  \right\rangle}

\newcommand{\dagg}[1]{#1^\dagger}
\newcommand{\tr}{\mathrm{tr}}

\begin{document}

\title{High fidelity feed-back assisted parity measurement in circuit QED}

\author{L. Tornberg and G. Johansson}
\affiliation{Department of Microtechnology and Nanoscience - MC2,Chalmers University of Technology, SE-41296
Gothenburg, Sweden}

\date{\today}
\begin{abstract}
We analyze a two qubit parity measurement based on dispersive read-out in circuit quantum electrodynamics. The back-action on the qubits has two qualitatively different contributions. One is an unavoidable dephasing in one of the parity subspaces, arising during the transient time of switching on the measurement. The other part is a stochastic rotation of the phase in the same subspace, which persists during the whole measurement. The latter can be determined from the full measurement record, using the method of state estimation. Our main result is that the outcome of this phase determination process is {\em independent} of the initial state in the state estimation procedure.
The procedure can thus be used in a measurement situation, where the initial state is unknown. We discuss how this feed-back method can be used to achieve a high fidelity parity measurement for realistic values of the cavity-qubit coupling strength. Finally, we discuss the robustness of the feed-back procedure towards errors in the measurement record.\end{abstract}

\pacs{PACS numbers: 42.50.Dv,  03.67.Pp, 02.30.Yy, 42.50.Pq}

\maketitle
\section{Introduction}
In circuit quantum electrodynamics (cQED) \cite{BlaisPRA69, WallraffNature431}, superconducting qubits \cite{WendinShumeiko2005, YouPhysToday2005, ClarkeNature2008}
are coupled to a microwave cavity, which allows for high fidelity qubit state control \cite{ChowPRL2009} and measurement \cite{DiCarloNature2009, ChowArXiv,FilippPRL2009}, two ingredients required to achieve a scalable quantum information architecture \cite{DiVincenzo00}. In the  dispersive regime, where the cavity and qubits are sufficiently detuned \cite{BlaisPRA69}, the state of the qubits shifts the resonance frequency of the cavity which can be detected by a homodyne measurement of the field state. This realizes a joint read-out of the state of the qubits and allows for measurement of single, as well as, multi-qubit operators \cite{DiCarloNature2009, ChowArXiv,FilippPRL2009}. Such a resource allows for measurement of the state parity which in turn enables deterministic entanglement through measurement \cite{BeenakkerPRL2004, EngelScience2005, Lalumiere} and
implementation of error correction protocols \cite{NielsenChuang, Ahn, Sarovar}. A true two-qubit parity measurement projects the state of the system on either the even parity subspace spanned by $\ket{gg}$ and $\ket{ee}$, or on the odd parity subspace spanned by $\ket{ge}$ and $\ket{eg}$, simultaneously allowing the observer to infer which of the two projections has occurred. The measurement is however not allowed to give information about the nature of the states within the respective subspaces as this would destroy the quantum superposition of the post-measurement state \cite{Neumann}. \\\\
In the dispersive read-out, the entanglement between qubits and field allows the observer to infer the parity of the qubit state. As analyzed by Lalumi\`ere et. al. in Ref.~[\onlinecite{Lalumiere}], the same entanglement gives rise to an unavoidable back-action causing dephasing within one of the parity subspaces. This severely limits the fidelity of the measurement and causes the post-measurement state to be mixed and thus limited for further use in a computational context. \\\\
In this paper, we show that the full measurement record from the homodyne current together with quantum feedback based on state estimation \cite{DohertyPRA62, DohertyPRA60}, can be
used to significantly reduce the unwanted dephasing caused by the measurement. This possibility was previously analyzed in a different setup where the homodyne detection could be regarded as
a strong projective measurement \cite{BarrettPRA71}. 
For circuit QED, a weak measurement is more realistic, and in the present work 
we extend the analysis to this situation, using a quantum trajectories approach \cite{WisemanPRA47}.
\section{Dispersive read-out as a parity measurement}\label{sec:p_measurement}
\begin{figure}[!ht]
\begin{center}
\includegraphics[width=\columnwidth]{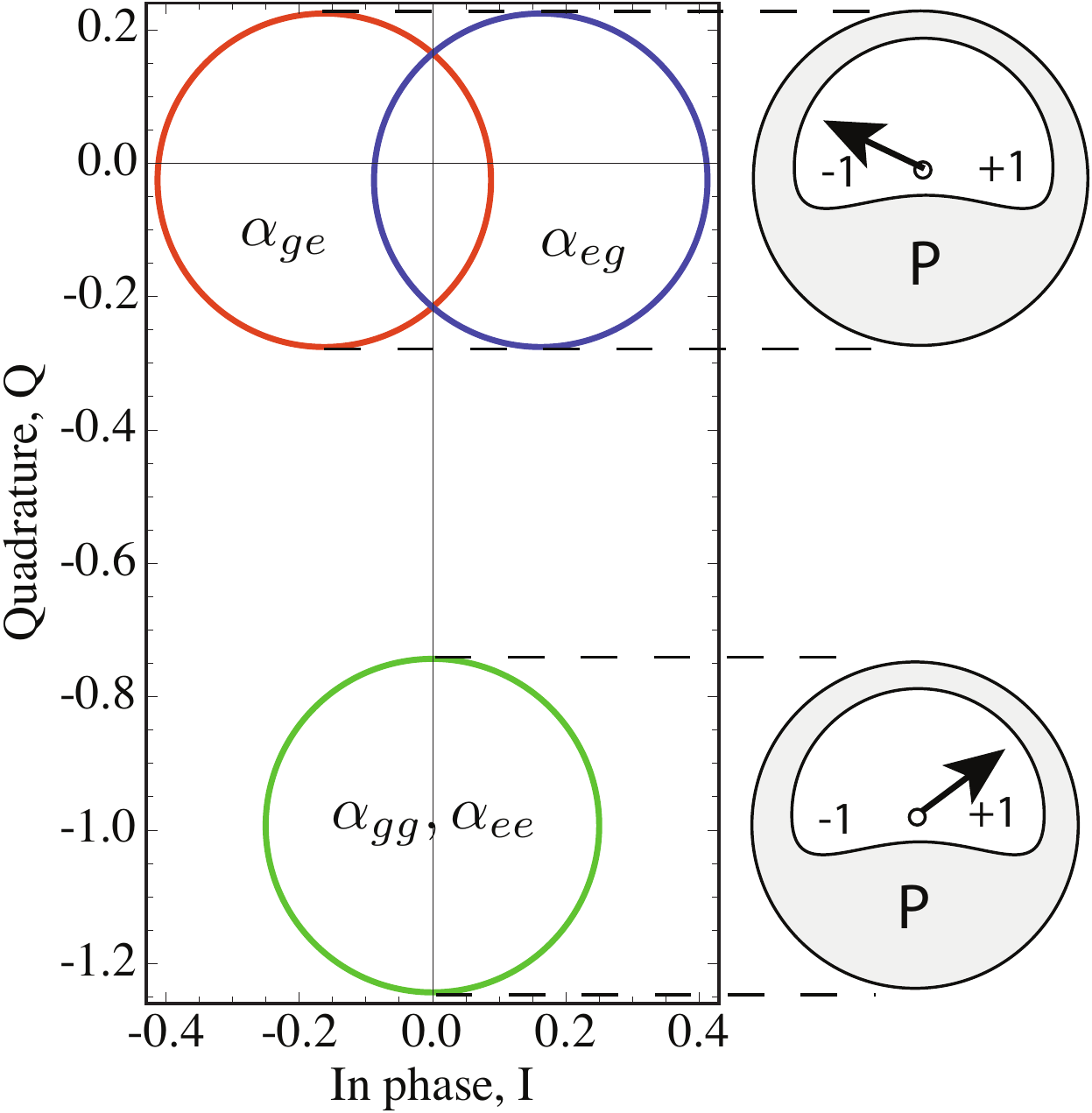}
\caption{Phase space illustration of the stationary field states given in \Eq{eq:amplitudes} with parameters  $\Delta_r = 0$, $\epsilon_m = 0.5 \kappa$, $\chi_1 = -\chi_2 = 1.5 \kappa$. By choosing the relative phase of the local oscillator to be $\phi = \pi/2$, the homodyne read-out measures the projection of the field on the $Q$-quadrature allowing the observer to infer the parity of the qubits. } \label{fig:fieldamplitudes}
\end{center}
\end{figure}

In this section, we lay out the essential idea behind the realization of a two qubit parity measurement using dispersive read-out in circuit QED \cite{BlaisPRA69, WallraffNature431}. 
For two qubits, the objective of a parity measurement is to distinguish between measurement results corresponding to states belonging to the two othogonal subspaces $\mathcal{H}_+ = \text{span}(\ket{gg},\ket{ee})$ and $\mathcal{H}_- = \text{span}(\ket{ge},\ket{eg})$ with corresponding parities $P_+ = 1$ and $P_- = -1$. Furthermore, such a measurement is not allowed not distinguish between states within the respective subspaces since this would destroy superpositions of states in either $\mathcal{H}_+$ or $\mathcal{H}_-$. 
We consider a cQED system consisting of two superconducting charge qubits coupled to a single field mode of a superconducting stripline resonator \cite{BlaisPRA69, WallraffNature431}. Neglecting the influence of higher qubit states, the system is well described by the Tavis-Cummings Hamiltonian \cite{TavisPhysRev170}
\bea
H &=& \omega_r \dagg{a}a +
\sum_{j=1}^2\frac{\omega_{qj}}{2}\sigma_z^{(j)} +
\sum_{j=1}^2 g_j(a\sigma_-^{(j)} + \dagg{a}\sigma_+^{(j)}) \nonumber \\
&+& 
(a\epsilon_m^\ast e^{i\omega_mt} + \dagg{a}\epsilon_m e^{-i\omega_mt}),
\eea
where we have set $\hbar = 1$, $\omega_r$ is the resonance frequency of the cavity mode and $\epsilon_m$ is the drive amplitude of the measurement signal tuned to frequency $\omega_m$. The level splitting of qubit $j$ is given by $\omega_{qj}$ and $g_j$ is the coupling between the corresponding qubit and the cavity. To realize a qubit measurement, the system is operated in the dispersive regime where the cavity and qubit frequencies are far detuned, $\lambda_j = g_j/|\Delta_j| \ll 1$ where $\Delta_j = \omega_r - \omega_{qj}$. In this limit the system is well described by the second order effective Hamiltonian \cite{BlaisPRA69}
\bea\label{eq:H}
H_\text{eff} &=& \Delta_r \dagg{a}a + \sum_{j=1}^2 \chi_j\sigma_z^{(j)}\dagg{a}a + \frac{\omega_{qj} + \chi_j}{2}\sigma_z^{(j)} \\
&+& (a\epsilon^\ast_m + \dagg{a} \epsilon_m ) + 
J_{12}(\sigma_-^{(1)}\sigma_+^{(2)} +  \sigma_+^{(1)}\sigma_-^{(2)}) ,\nonumber
\eea
written in the frame where the cavity degrees of freedom rotate at the frequency $\omega_m$ such that  $\Delta_r = \omega_r - \omega_m$. The residual coupling between cavity and qubit $j$ is described by  $\chi_j = g_j^2/\Delta_j$ and  $J_{12} = g_1g_2(\Delta_1 + \Delta_2)/2\Delta_1\Delta_2$ is the qubit-qubit coupling mediated by virtual photons \cite{BlaisPRA75}. \\\\
In the dispersive readout of cQED, the joint state of the qubits is inferred from the homodyne signal coming from the transmitted microwaves through the cavity. Due to the coupling between cavity and qubits, the phase and amplitude of the transmitted microwaves will depend on the state of the qubits $\ket{ij}$, $i,j = g,e$ with the field evolving into a coherent state $\ket{\alpha_{ij}}$ with amplitude $\alpha_{ij}$ which obeys the differential equation \cite{GambettaPRA74}
\be\label{eq:amplitudes}
\dot{\alpha}_{ij}(t) = -i\epsilon_m - i(\Delta_r + \chi_{ij})\alpha_{ij}(t) - \frac{\kappa}{2}\alpha_{ij}(t).
\ee
Here, $\chi_{ij} = \bra{ij}\chi_1 \sigma_z^{(1)} + \chi_2 \sigma_z^{(2)}\ket{ij}$ is the coupling between the state $\ket{ij}$ and the cavity. In this case the four different cavity states can be interpreted as pointer states of the measurement, where the state of the two qubits can be inferred from the state of the cavity field. To realize a parity measurement we must make sure that the read-out distinguishes between the two sets of field states $\{\alpha_{gg},\alpha_{ee}\}$ and $\{\alpha_{ge},\alpha_{eg}\}$ but not between the amplitudes within the sets. By choosing the couplings $g_1 = g_2$ and detunings $\Delta_1 = -\Delta_2$, the four different amplitudes in \Eq{eq:amplitudes} satisfy
\bea\label{eq:amplitudes_parity}
\alpha_{gg}(t) &=& \alpha_{ee}(t),\nonumber \\
\text{Re}(\alpha_{ge}(t)) &=& -\text{Re}(\alpha_{eg}(t)) \nonumber \\
\text{Im}(\alpha_{ge}(t)) &=& \text{Im}(\alpha_{eg}(t)) .
\eea
which can be seen in \Fig{fig:fieldamplitudes}, were the quadrature $Q = \text{Im}(\alpha_{ij}(t))$ is plotted against the in-phase component of the field $I = \text{Re}(\alpha_{ij}(t))$. 
In a homodyne measurement the field of the cavity is mixed with a local oscillator with a relative phase $\phi$ defining the measured quadrature. By choosing $\phi = \pi/2$, the $Q$-quadrature is measured as indicated in \Fig{fig:fieldamplitudes}. In this case, the observer can distinguish between states of even and odd parity. The signal will however not contain information about the phase between states $\{\ket{ge},\ket{eg}\}$ and $\{\ket{gg},\ket{ee}\}$ as required by a parity measurement.  Since the measurement eigenstates are not eigenstates of the qubit-qubit coupling (last term in \Eq{eq:H}) we choose $\Delta_1 = -\Delta_2$ to make this vanish. 
\section{The model}\label{sec:themodel}
\subsection{Effective two qubit stochastic master equation}
The appropriate equation of motion to describe the evolution of the system conditioned on continuous homodyne detection is given by the stochastic master equation (SME) \cite{WisemanPRA47}
\be\label{eq:SME_combined}
d\rho_c = \mathcal{L}_\text{tot}\rho_c dt  +  \sqrt{\kappa\eta}\mathcal{M}[a e^{-\phi}]\rho_c dW(t) ,
\ee
where $\mathcal{L}_\text{tot}\rho_c$ is given by 
\bea\label{eq:master_start}
\mathcal{L}_\text{tot}\rho_c &=& -i[H_\text{eff},\rho_c]   
+ \sum_{j}\gamma_{1j}\mathcal{D}[\sigma_-^{(j)}]\rho_c + \frac{\gamma_{\phi j}}{2}\mathcal{D}[\sigma_z^{(j)}] \rho_c\nonumber \\
&+&\kappa\mathcal{D}[a]\rho_c +  \kappa \mathcal{D}[ \sum_{j}\ \lambda_j\sigma_-^{(j)}]\rho_c,
\eea
where $\mathcal{D}[c]\rho = c\rho\dagg{c} - 1/2(\dagg{c}c \rho + \rho \dagg{c}c)$ is a dissipation super operator in Lindblad form \cite{Lindblad} with the pure relaxation and dephasing rates of qubit $j$ given by $\gamma_{1j}$ and $\gamma_{\phi j}$ respectively. The cavity damping rate is given by $\kappa$ and the last term of \Eq{eq:master_start} describes the effect of Purcell damping \cite{PurcellPhysRev69}. The measurement back-action on the system is described by the measurement super-operator  
\be
\mathcal{M}[c] \rho_c = c\rho_c + \rho_c \dagg{c} - \langle c + \dagg{c}\rangle \rho_c,
\ee
where $\langle \cdot \rangle = \tr(\cdot \rho_c)$ and $dW(t)$ is defined as a Wiener increment completely characterized by its mean and variance \cite{Klebaner}
\bea
\text{E}[dW(t)] &=& 0, \nonumber \\
\text{E}[dW(t)^2] &=& dt.
\eea
Here, $\text{E}[\cdot]$ denotes the ensemble average taken over different realizations of the noise process $W(t)$. The homodyne current is given by 
\be
j(t)dt = \sqrt{\kappa\eta} \langle a e^{-\phi} + \dagg{a }e^{\phi}\rangle dt  + dW(t),
\ee 
where $\eta$ is the efficiency at which the photons are detected. \\\\
The aim of this section is to trace out the cavity degrees of freedom from \Eq{eq:master_start} to obtain an effective master equation for the two qubits only. As shown in Refs. [\onlinecite{Lalumiere, GambettaPRA77}] this can be done in the limit $\gamma_{1j} \ll \kappa$, which is easily satisfied with current Purcell limited qubits \cite{HouckPRL101}. In a frame of reference given by the transformation 
\be\label{eq:P}
\mathbf{P} = \sum_{i,j = g,e}\Pi_{ij}D[\alpha_{ij}],
\ee
the photon population in the cavity is essentially zero, which allows one to trace out the cavity degrees of freedom. Here, $D[\alpha] = \exp(\alpha \dagg{a} - \alpha^\ast a)$ is the displacement operator of the field \cite{GerryKnight} and $\Pi_{ij} = \ket{ij}\bra{ij}$ are projection operators onto the respective basis state of the two qubit Hilbert space.  This gives an effective master equation for the qubit degrees of freedom 
\bea\label{eq:reduced_master}
d\rho_c &=& -i\left[\sum_j \frac{\omega_{qj} + \chi_j}{2}\sigma_z^{(j)}, \rho_c\right]dt  \nonumber \\
&+&\left(\sum_{j}\gamma_{1j}\mathcal{D}[\sigma_-^{(j)}]\ + 
\frac{\gamma_{\phi j}}{2}\mathcal{D}[\sigma_z^{(j)}] + 
\kappa \mathcal{D}[ \sum_{j}\ \lambda_j\sigma_-^{(j)}] \right) \rho dt \nonumber \\
&+& \sum_{ij,kl} \chi_{kl,ij}(\text{Im}(\alpha_{ij}^\ast\alpha_{kl}) + i \text{Re}(\alpha_{ij}^\ast\alpha_{kl})) \Pi_{ij}\rho_c\Pi_{kl} dt \nonumber \\
&+& \sqrt{\kappa \eta} \mathcal{M}[\Pi_\alpha e^{-i\phi}]\rho_c dW(t),
\eea
where 
$\chi_{ij,kl} = \chi_{ij} - \chi_{kl}$. The fifth term represents the dephasing induced by the measurement and the sixth term gives the AC Stark shift. The measurement operator $\Pi_\alpha$ is given by
\bea
\Pi_\alpha &=&  \sum_{ij}\alpha_{ij}\Pi_{ij} \\
&=&\frac{1}{4}\left[\gamma_{zz} \sigma_z^{(1)}\sigma_z^{(2)} + \gamma_{z1}\sigma_z^{(1)} + \gamma_{1z}\sigma_z^{(2)} + \gamma_{11} \right], \nonumber
\eea
with the amplitudes
\bea
\gamma_{zz} &=& \alpha_{gg} - \alpha_{ge} - \alpha_{eg} + \alpha_{ee}, \nonumber \\
\gamma_{z1} &=& -\alpha_{gg} - \alpha_{ge} + \alpha_{eg} + \alpha_{ee}, \nonumber \\
\gamma_{1z} &=& -\alpha_{gg} + \alpha_{ge} - \alpha_{eg} + \alpha_{ee}, \nonumber \\
\gamma_{11} &=& \alpha_{gg} + \alpha_{ge} + \alpha_{eg} + \alpha_{ee} .
\eea
By choosing system parameters appropriately, the homodyne measurement along with single qubit rotations can thus be tuned to measure single, as well as two qubit operators as discussed in Refs. [\onlinecite{Lalumiere, DiCarloNature2009, FilippPRL2009, ChowArXiv}]. The current can be expressed in terms of qubit operators as 
\be\label{eq:current_qb}
j(t)dt =\sqrt{\kappa\eta}\langle \Pi_\alpha e^{-i\phi} + \dagg{\Pi_\alpha} e^{i\phi}\rangle dt  + dW(t),
\ee
which in the case of $\phi = \pi/2$ reduces to 
\be
j_\text{ij}(t)dt  = 2\sqrt{\kappa\eta}\text{Im}(\alpha_{ij})dt  + dW(t),
\ee
for the basis states $\ket{ij}$. With our choice of parameters, $\text{Im}(\alpha_{gg})= \text{Im}(\alpha_{ee})$ and  $\text{Im}(\alpha_{ge})= \text{Im}(\alpha_{eg})$ such that the measurement does not distinguish between states within $\mathcal{H_+}$ and $\mathcal{H_-}$ as required for a parity measurement. 
\section{Measurement induced dephasing}\label{sec:dephasing}
\begin{figure}[!ht]
\begin{center}
\includegraphics[width=\columnwidth]{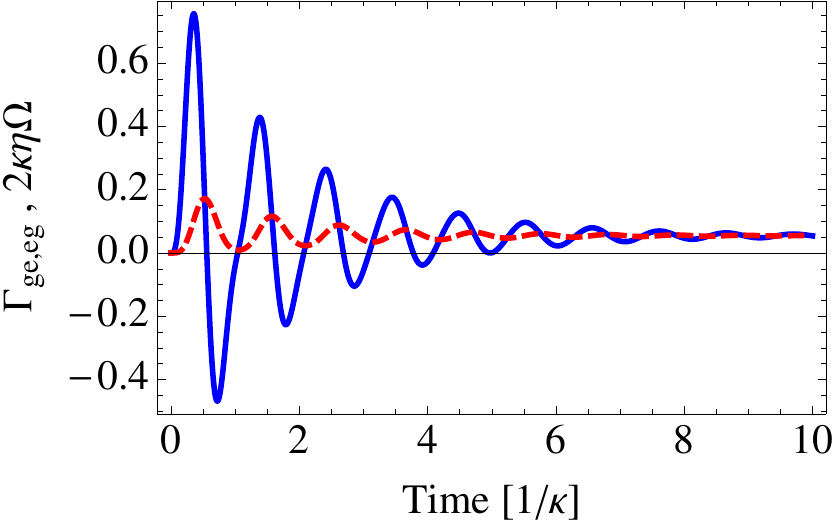}
\caption{(Color online) The dephasing rates $\Gamma_{ge,eg}$ (solid blue line) and $2\kappa\eta \Omega$ (dashed red line). Initially, the system is prone to dephasing wth $\Gamma_{ge,eg} > 2\kappa\eta \Omega$.  This reflects the fact that initial fluctuations in the photon number can not be captured by the homodyne measurement. As $t\to\infty$, $\Gamma_{ge,eg} \to 2\kappa\eta \Omega$ with all the unwanted back action given by a stochastic phase between the basis states $\ket{ge}$ and $\ket{eg}$. The parameters are $\epsilon_m = \kappa$, $\chi_1 = -\chi_2 = 3 \kappa$ and $\eta = 1$. 
 } 
 \label{fig:deph_rate_comp}
\end{center}
\end{figure}

With the field amplitudes given in \Eq{eq:amplitudes_parity} and the LO-phase given by $\phi = \pi/2$ the measurement operator can be rewritten as
\bea\label{eq:measop_parity}
\mathcal{M}[-i\Pi_\alpha] \rho_c =
\frac{\beta}{2} \mathcal{M}[ \Pi_+ - \Pi_-] \rho_c - i\text{Re}(\alpha_{ge})[\Pi_{ge} - \Pi_{eg},\rho_c], \nonumber \\
\eea
where 
\bea
\Pi_+ &=& \ket{gg}\bra{gg}  + \ket{ee}\bra{ee}, \nonumber \\
\Pi_- &=& \ket{ge}\bra{ge}  + \ket{eg}\bra{eg},
\eea
are projection operators onto $\mathcal{H_+}$ and $\mathcal{H_-}$ and $\beta$ is the difference field 
\be
\beta =  \text{Im}(\alpha_{gg}) - \text{Im}(\alpha_{ge}).
\ee 
The first term of \Eq{eq:measop_parity} is associated with information gain and simply expresses the fact that the parity measurement tends to localize an initial state in one of the subspaces $\mathcal{H}_+$ or $\mathcal{H}_-$ at a rate given by
\be
\Gamma_\text{m}(t) = \kappa\eta\beta(t)^2,
\ee
which can be taken as a definition of the measurement rate. The second term represents a stochastic phase between the states $\ket{ge}$ and $\ket{eg}$ accumulated during the measurement. This, together with the dephasing term $\propto \text{Im}(\alpha_{ij}^\ast\alpha_{kl})$ given in \Eq{eq:reduced_master}, represents the unwanted back-action of the measurement which we now analyze. \\\\
Because of the non-linear nature of conditional state evolution, it is not possible to analytically solve \Eq{eq:reduced_master} in the general case. However, in the case when the state has been projected onto $\mathcal{H}_\pm$ the non-linear part vanishes and \Eq{eq:reduced_master} reduces to 
\bea\label{eq:stoch_diff}
d\rho_c(t) &=& \frac{ \Gamma_{ge,eg} }{2}\mathcal{D}[\Sigma_z]\rho_c(t) dt  
- i\sqrt{\kappa \eta \Omega} \mathcal{K}\rho_c(t) dW(t), \nonumber \\
\eea
where $\Sigma_z = \Pi_{ge} - \Pi_{eg}$, $\mathcal{K}\rho_c =  [\Sigma_z,\rho]$, $\Gamma_{ge,eg}=4\chi  \text{Im}(\alpha_{ge}^\ast\alpha_{eg})$ and $\Omega = \text{Re}^2(\alpha_{ge})$. Since the focus of the analysis is on the back action of the measurement we have neglected the effect of pure relaxation and dephasing as well as Purcell damping. Furthermore, we disregard the deterministic rotation of the phase $\propto \text{Re}(\alpha_{ij}^\ast\alpha_{kl})$ given in \Eq{eq:reduced_master} since this can be undone regardless of the measurement outcome. \Eq{eq:stoch_diff} can be solved analytically \cite{Klebaner}, yielding 
\bea\label{eq:sdiff_sol}
\rho_c(t) &=& \exp\Big[  \frac{1}{2}\int_0^t \Gamma_{ge,eg}(s)  \mathcal{D}[\Sigma_z]  
 + \kappa \eta \Omega(s) \mathcal{K} ^2 ds \nonumber \\  
&-& i\int_0^t \sqrt{\kappa \eta \Omega(s)}dW(s) \mathcal{K} \Big]\rho(0) \nonumber \\
& = & \exp\Big[\frac{1}{2}  \int_0^t \big( \Gamma_{ge,eg}(s)  - 2 \kappa \eta \Omega(s)\big)\mathcal{D}[\Sigma_z]  ds \nonumber \\  
&-& i\int_0^t \sqrt{\kappa \eta \Omega(s)}dW(s) \mathcal{K} \Big]\rho(0),
\eea
where we have used the fact that $\mathcal{K}^2 \rho = -2\mathcal{D}[\Sigma_z]\rho$. Given this, we see two effects of the measurement. First, the fourth line in \Eq{eq:sdiff_sol} gives the dephasing rate $\Gamma_c$ for a single trajectory 
\be\label{eq:dephrate_c}
\Gamma_c(t) =  \Gamma_{ge,eg}(t)  - 2 \kappa \eta \Omega(t).
\ee
The second effect is given by the last term in \Eq{eq:sdiff_sol} which gives rise to a stochastic phase between $\ket{ge}$ and $\ket{eg}$. Since this phase varies between different measurements the ensemble averaged state is mixed with the corresponding dephasing rate given by $\Gamma_{ge,eg}$. As a measure of the read-out fidelity we take the ratio between the measurement and dephasing rate as $t\to \infty$
\be\label{eq:ratio_nofb}
\frac{\Gamma_{ge,eg}}{\Gamma_\text{m}}_{|t\to\infty} = \frac{\kappa^2}{8\eta \chi^2},
\ee
which shows that the dispersive read-out only works as a perfect parity measurement in the large coupling limit $\chi \gg \kappa$. This limit may however be experimentally hard to reach \cite{Lalumiere} and it is therefore desirable to seek alternative methods to improve upon this ratio. Considering the same ratio but for a single trajectory gives the result  
\be\label{eq:ratio}
\frac{\Gamma_c}{\Gamma_\text{m}}_{|t\to\infty} = \frac{(1-\eta)\kappa^2}{8\eta \chi^2},
\ee
which vanishes in the case of a perfect measurement $\eta = 1$. This can be understood by looking at the two terms in \Eq{eq:dephrate_c}. In \Fig{fig:deph_rate_comp}, we see that for a perfect measurement, $\Gamma_c \to 0$ as $t\to \infty$, which is due to the fact that the measurement current contains information about the photon fluctuations. As $t \to \infty $, the observer can track the photon fluctuations by looking at the homodyne current such that no information is lost into the environment. Given this information, the observer can in principle undo the stochastic phase with a resulting pure ensemble averaged state. From \Fig{fig:deph_rate_comp}, we also see that, for the dispersive measurement, there is initial dephasing in the system not caught when taking the limit $t\to\infty$. This is given by  
\be\label{eq:dephasingtransient}
\lim_{\tau \to \infty } \int_0^\tau \Gamma_c(s) ds = \frac{128\epsilon^2\chi^2}{(\kappa^2 + 16\chi^2)^2}, \qquad \eta = 1,
\ee
which simply reflects the fact that the homodyne measurement is not able to record information about the initial photon fluctuations due to the finite response time of the cavity. Because of this, inevitable dephasing occurs with the final state being mixed even for a single trajectory.
The dephasing in \Eq{eq:dephasingtransient} is a decreasing function of $\chi$ since the number of photons in the cavity interacting with the qubits decreases as the coupling to the qubits shifts the cavity frequency. Moreover, the dephasing decreases with increasing $\kappa$. This is simply due to the fact that a faster response time of the cavity allows for a more efficient measurement of the initial photon fluctuations. \\\\
From \Eq{eq:ratio}, we note that the measurement fidelity of a single trajectory is far better than that given for the ensemble averaged state in \Eq{eq:ratio_nofb}. This is because the stochastic phase does not contribute to the dephasing for a single trajectory. From this it is clear that the purity of the ensemble averaged state can be drastically increased if the observer is able to undo the stochastic phase after each measurement shot. In this case the dispersive read-out approaches a true parity measurement even in the case of moderate coupling between qubits and cavity. In the next section we present such a scheme based on state estimation and feed-back. We show that this can be done without knowledge of the initial state. 
\section{Increasing measurement fidelity by state estimation}\label{sec:feedback}
Information about the stochastic phase discussed in \Sec{sec:dephasing} can not be extracted from the measurement current directly. We may however use the recorded current to extract the noise process $W(t)$ for a given trajectory. This can be used as an input to \Eq{eq:reduced_master} to calculate an estimate $\tilde{\rho}_c(t)$ of the state \cite{DohertyPRA62, DohertyPRA60}.
Unfortunately, this procedure requires knowledge of the initial state $\rho_c(0)$ which is typically unknown in a measurement situation. We now show that no information about the initial state is needed if we only are interested in the accumulated stochastic phase. \\\\
Given an unknown initial state, the observer will make a wrong estimate $d\tilde{W}(t)$ of the true noise process
\bea
d\tilde{W}(t) &=& dW(t) \\
&+&\sqrt{\kappa\eta}\tr\left[ \left(\Pi_\alpha e^{-i\phi} + \dagg{\Pi_\alpha} e^{i\phi}\right)(\rho_c(t) - \tilde{\rho}_c(t))\right] dt, \nonumber
\eea
where $\rho_c(t)$ is the true state of the system and $\tilde{\rho}_c(t)$ the guessed estimate. We write the true state as a mix of odd and even parity states
\be\label{eq:rho_init}
\rho_c(t) = p(t)\rho_+(t)+   (1-p(t))\rho_-(t)
\ee
where $\rho_+(t) \in \mathcal{H_+}$ and $\rho_-(t) \in \mathcal{H_-}$. The probability $p(t)\in\{0,1\}$ describes the conditional evolution towards one of the respective subspaces. We choose the estimated state to initially reside in the $\mathcal{H}_-$ subspace such that the subsequent state can be written
\be
\tilde{\rho}_c(t) = \tilde{\rho}_-(t). 
\ee
Considering the measurement part of \Eq{eq:reduced_master} only, we can write three coupled equations for the $\ket{ge}\bra{eg}$ matrix element of $\rho_c$, $\tilde{\rho}_c$ and $p(t)$
\bea
d\rho_{ge,eg} &=& -2\sqrt{\kappa \eta}\big(\beta p + i \text{Re}(\alpha_{ge})\big) \rho_{ge,eg}dW(t),\nonumber\\
d\tilde{\rho}_{ge,eg} &=& -i 2\sqrt{\kappa \eta}\big(2 \sqrt{\kappa \eta} \beta p dt  + dW(t)\big)\text{Re}(\alpha_{ge})  \tilde{\rho}_{ge,eg}, \nonumber \\
dp(t) &=&  -2\sqrt{\kappa \eta}\beta p(1-p)dW(t).
\eea
Given the solution for $p(t)$, we can calculate the phases $\phi = \arg(\rho_{ge,eg})$ and $\tilde{\phi} = \arg(\tilde{\rho}_{ge,eg})$ of the true and estimated state respectively
\bea\label{eq:mainres}
\phi(t) &=& \tilde{\phi}(t) = -4 \kappa \eta \int_0^t  \beta(s)\text{Re}(\alpha_{ge}(s)) p(s) dt \nonumber \\
&-& 2\sqrt{\kappa \eta } \int_0^t \text{Re}(\alpha_{ge}(s)) dW(t).
\eea 
Hence, we see that the accumulated phase of the estimated state equals that of the true state! This is the main result of this paper. This is consistent with the result of Ref.~[\onlinecite{BarrettPRA71}], where a strong projective homodyne measurement is considered. This can be seen as the limiting case when $\kappa\to\infty$ of our analysis. Since the conditional evolution does not depend on whether the true state is a mix or superposition between even and odd parity states, the main result in \Eq{eq:mainres} holds equally well when the true state is an arbitrary superposition state. Hence, the method of undoing the accumulated phase can be applied not only in the context of e.g. entanglement generation as considered in Ref.~[\onlinecite{Lalumiere}] but also in a real measurement situation where the initial state is unknown. This will considerably reduce the unwanted dephasing casued by the measurement back action.  
\section{Results} 
\begin{figure}[!ht]
\begin{center}
\includegraphics[width=\columnwidth]{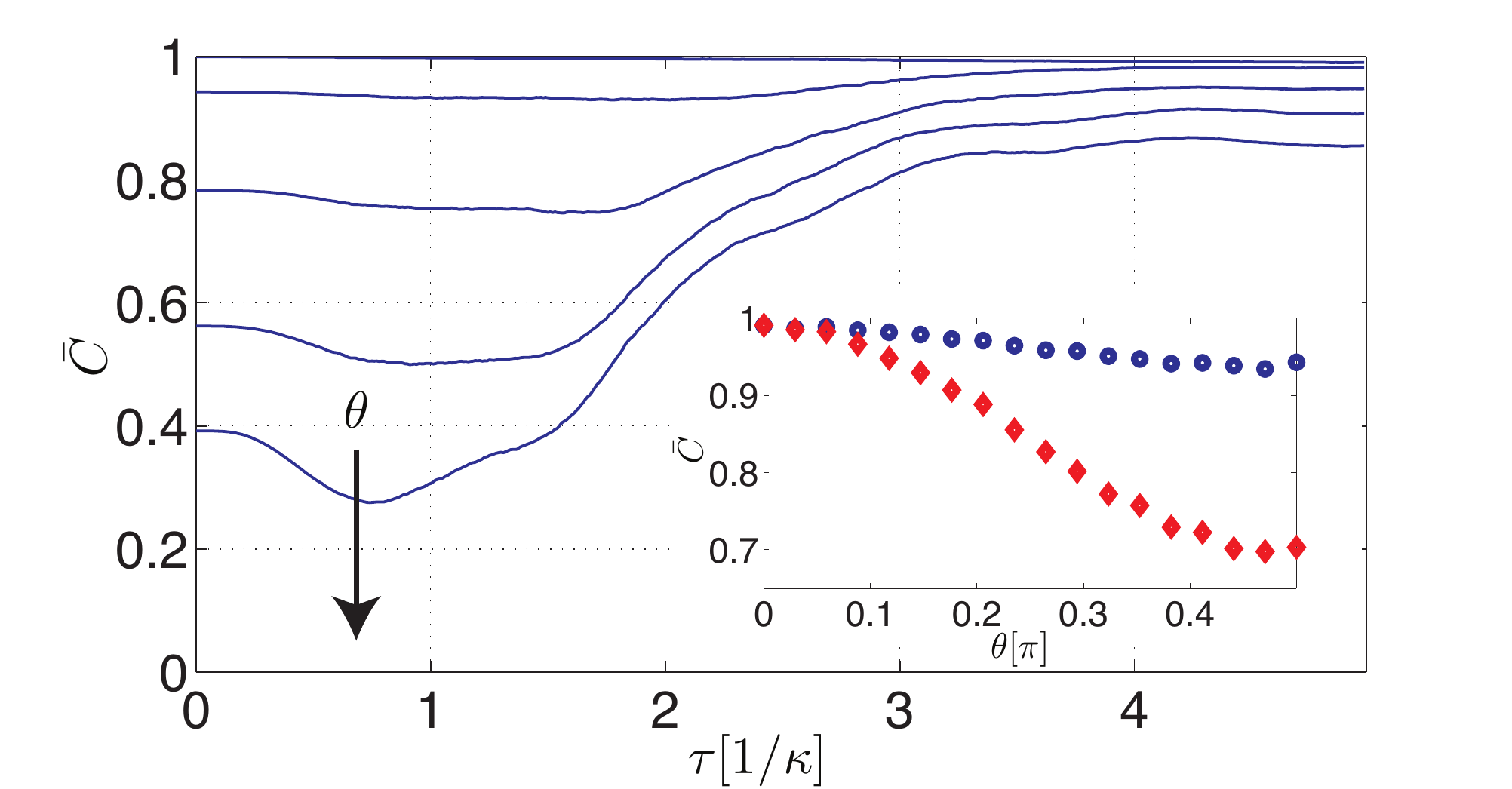}
\caption{(Color online) Average concurrence $\bar{C}$ as a function of time plotted for five initial states on the form given in \Eq{eq:rho_init}. The angle $\theta$ is in the range $\theta \in \{ 0, \pi/4 \}$ (increasing from top to bottom). The inset shows $\bar{C}$ at the final time of the measurement, with (blue $\circ$) and without (red $\diamond$) the phase subtracted. The parameters are given by  $\epsilon_m = \kappa$, $\chi_1 = -\chi_2 = 3 \kappa$, $g_1 = -g_2 = 100\kappa$, $\gamma_{1j} = \gamma_{\phi j} = 0$ and $\eta = 1$. } \label{fig:c}
\end{center}
\end{figure}

\begin{figure}[!ht]
\begin{center}
\includegraphics[width=\columnwidth]{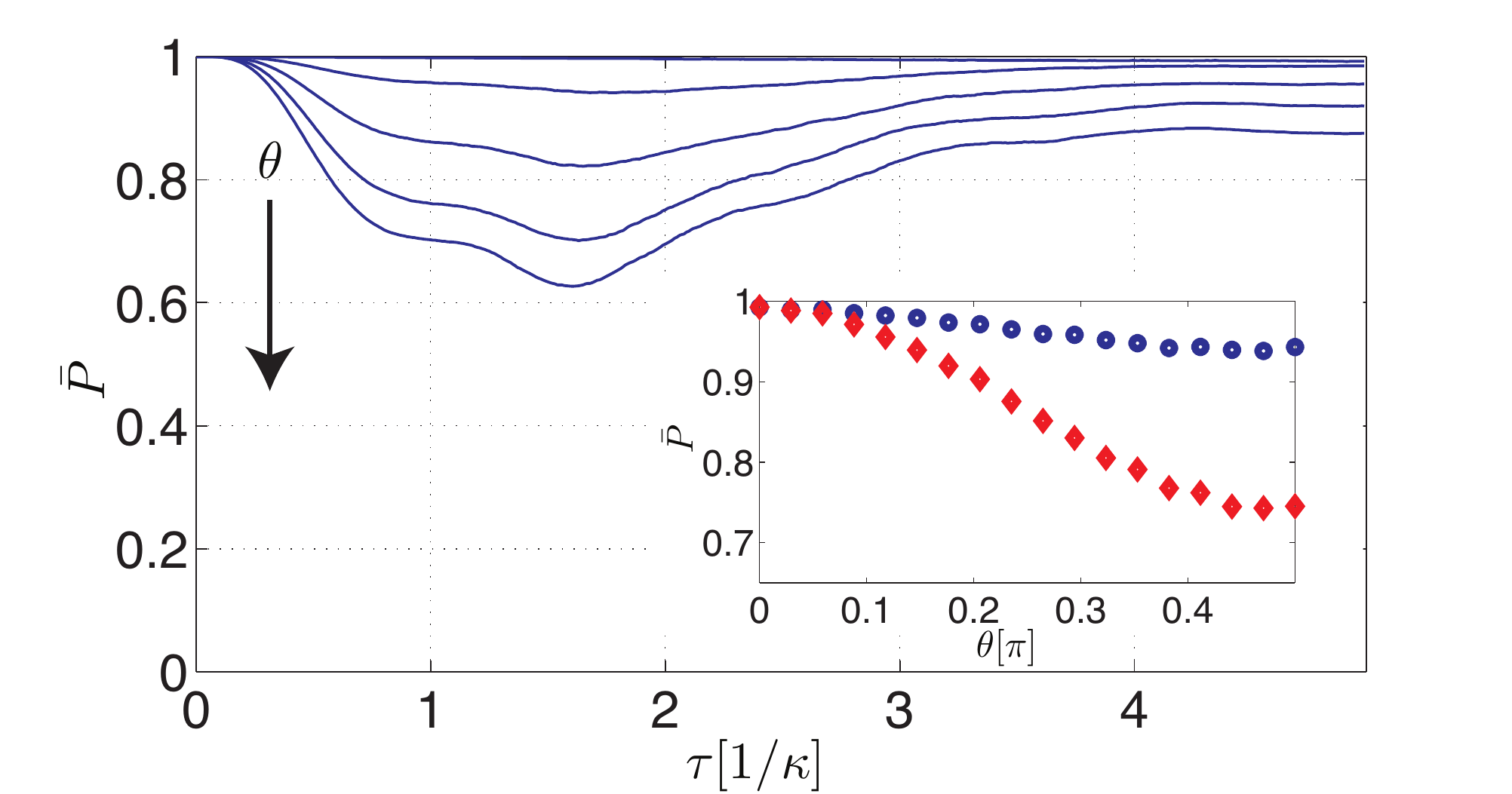}
\caption{(Color online) Average purity $\bar{P}$ as a function of time plotted for the same initial states as in \Fig{fig:c}. The inset shows $\bar{P}$ at the final time of the measurement, with (blue $\circ$) and without (red $\diamond$) the phase subtracted. The parameters are the same as in \Fig{fig:c}.} \label{fig:purity}
\end{center}
\end{figure}

\begin{figure}[!ht]
\begin{center}
\includegraphics[width=\columnwidth]{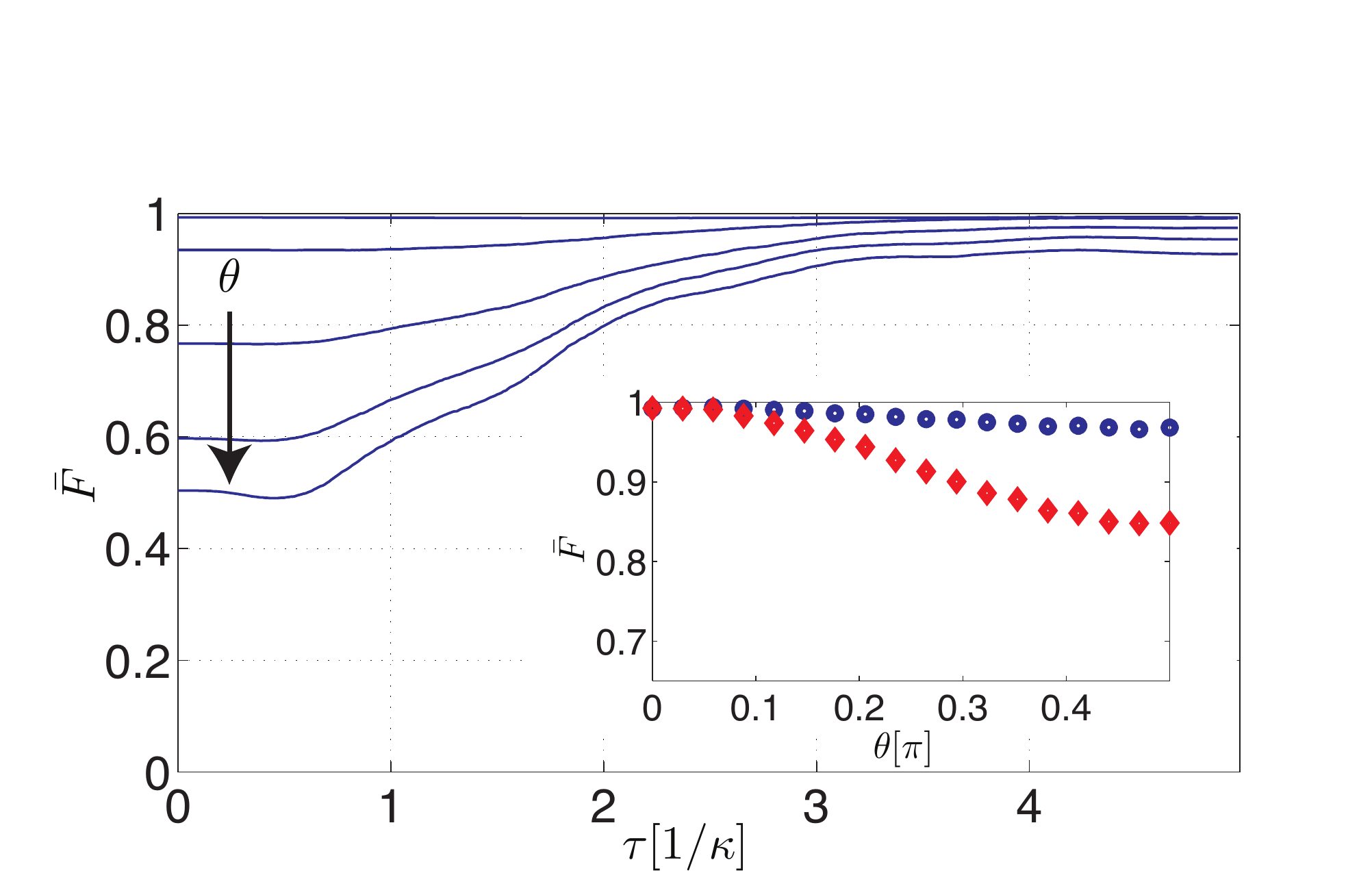}
\caption{(Color online) Average fidelity $\bar{F}$ as a function of time plotted for the same initial states as in \Fig{fig:c}. The inset shows $\bar{F}$ at the final time of the measurement, with (blue $\circ$) and without (red $\diamond$) the phase subtracted. The parameters are the same as in \Fig{fig:c}.} \label{fig:fidelity}
\end{center}
\end{figure}

To quantify how much the method of state estimation and subsequent subtraction of the accumulated phase increases the purity of the post-measurement state we perform numerical simulations of \Eq{eq:reduced_master}. This is necessary since the full dynamics of \Eq{eq:reduced_master} can not be solved analytically. To distinguish the odd and even parity states we use the integrated current
\be
s(t) = \int_0^t j(s)ds,
\ee
and assign the post measurement states to $\mathcal{H}_+$ or $\mathcal{H}_-$ if $s(t)$ satisfies $s(t) > s_{th}$ or $s(t) < s_{th}$ where $s_{th}$ is a threshold given by the average current $s_{th} = \sqrt{\kappa\eta}\int_0^ t (\text{Im}(\alpha_{gg}) + \text{Im}(\alpha_{ge}))ds $. To quantify the fidelity of the parity measurement and the estimation scheme we define the average concurrence $\bar{C} = C(\text{E}[\rho_c])$, purity $\bar{P} = \tr\big((\text{E}[\rho_c])^2\big)$ and fidelity
\be
\bar{F} = \frac{ n_+\bra{\psi_+}\text{E}_+[\rho_c] \ket{\psi_+}   + n_-\bra{\psi_-}\text{E}_-[\rho_c] \ket{\psi_-}}{n_+ + n_-},
\ee
where $n_+/n_-$ is the number of trajectories assigned to measurement outcome $+/-$ and $\text{E}_\pm[\rho_c]$ is the ensemble average over the respective states. The states $\ket{\psi_\pm}$ are even and odd parity states with the initial state given by 
\be
\ket{\psi} = \cos(\theta)\ket{\psi_+} + \sin(\theta)\ket{\psi_-}.
\ee
The initial estimated state is correspondingly given by $\ket{\tilde{\psi}} = \ket{\tilde{\psi}_-}$ where $\ket{\tilde{\psi}_-}$ need not equal $\ket{\psi_-}$ \cite{stateref}.  We plot $\bar{C}$, $\bar{P}$ and $\bar{F}$ in \Fig{fig:c},  \Fig{fig:purity} and \Fig{fig:fidelity} for different values of $\theta$. In each figure the inset shows the value of the respective quantity at the final time with and without the subtraction of the accumulated phase. It is clear that the state estimation allows us to subtract the stochastic phase in each measurement run, resulting in a drastic improvement of  the state fidelity across all three measures. \\\\

\subsection{Imperfect measurement record}
\begin{figure}[!ht]
\begin{center}
\includegraphics[width=\columnwidth]{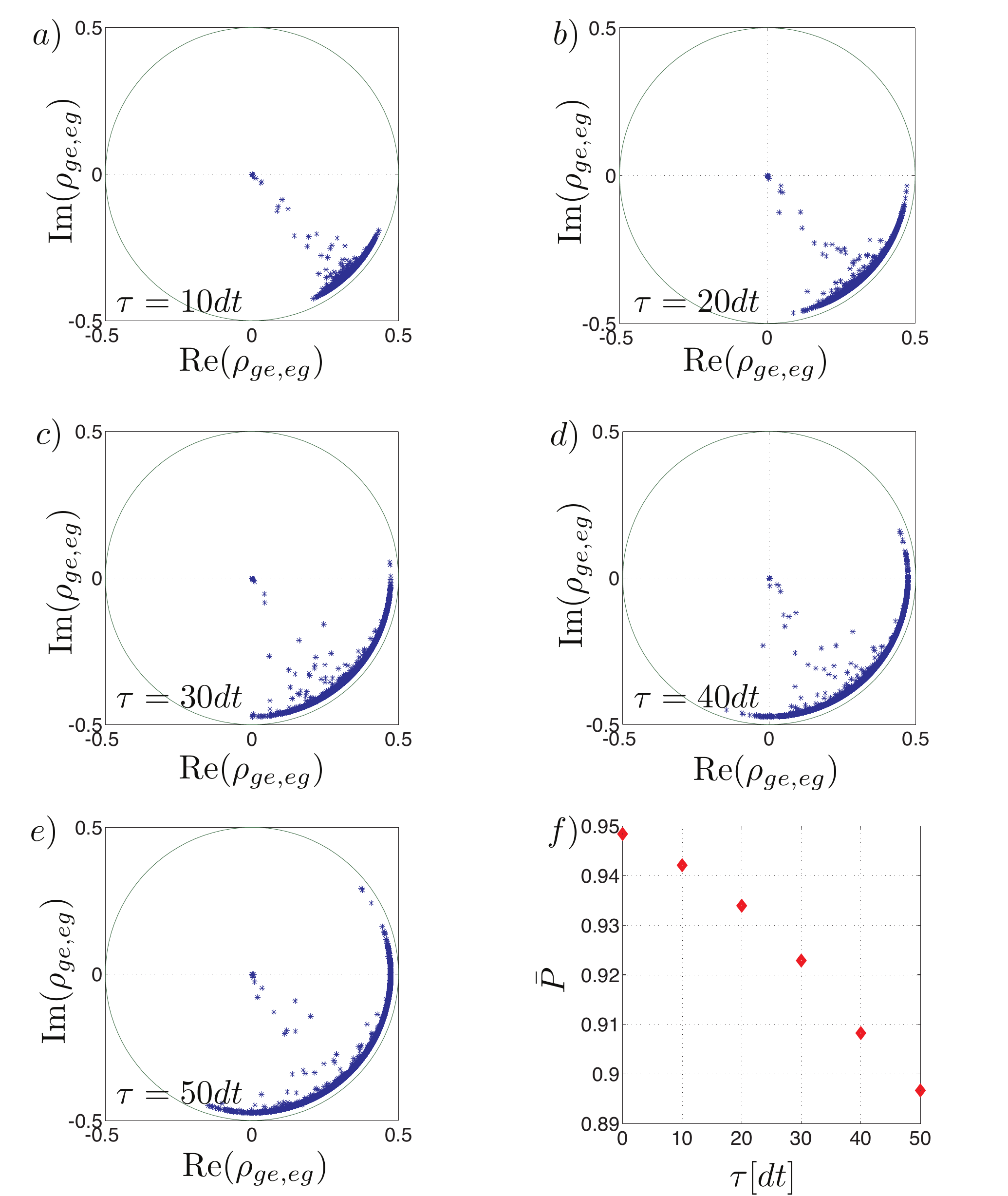}
\caption{(Color online) In a)-e), we plot $\text{Im}(\rho_{ge,eg})$ of the post measurement states against $\text{Re}(\rho_{ge,eg})$. The initial state is given by $\ket{\psi} = 1/\sqrt{2}(\ket{ge} + \exp(i\pi/4)\ket{eg})$. As $\tau$ increases the estimation of the accumulated phase becomes worse and the phase of the state is distributed around $\phi = -\pi/4$. In f), the average purity is plotted as function of $\tau$. Even for $\tau = 50dt$, the purity does not reach the corresponding value obtained in the case without state estimation and feedback. } \label{fig:tfilter}
\end{center}
\end{figure}

We note that our estimation scheme relies on the assumption that we can make a perfect record of the homodyne current, which is impossible in the case of a detector with limited bandwidth. We therefore consider the situation where the estimated state is based on the value of the low pass filtered current 
\be
j_\text{lp}(t) = \int_{t-\tau}^t j(s) ds,
\ee
where $\tau$ gives a measure of the deviation of $j_\text{lp}(t)$ from $j(t)$. In this case the value of the estimated phase will differ from the true value and the averaging over measurement results will reduce the purity of the state. In \Fig{fig:tfilter} we plot the imaginary and real part of $\rho_{ge,eg}$ of the post-measurement density matrix for each trajectory. The initial state is given by $\ket{\psi} = 1/\sqrt{2}(\ket{ge} + \exp(i\pi/4)\ket{eg})$.  
We see that the state estimation scheme successfully recovers the true phase of the initial state even in the case where we do not have access to a perfect measurement record. As $\tau$ increases the fidelity goes down as expected, but does not reach the value obtained without state estimation even for $\tau = 50dt$. We can understand this robustness from the observation that the values of $\rho_{ge,eg}$ are distributed on a circle arc which make the decrease of purity quadratic in the phase uncertainty. This can also be seen in the initial slope in \Fig{fig:tfilter}.\\\\
\section{Conclusion}\label{sec:conclusion}
In conclusion, we have performed an analysis of a two qubit parity measurement based on dispersive read-out in cQED. In particular, we analyzed the the back-action on the two qubits and found two qualitatively different contributions. One is an unavoidable dephasing in one of the parity subspaces, arising during the transient time of switching on the measurement. This dephasing occurs in the process of entangling the state of the driven cavity with the two qubits. The other part is a stochastic rotation of the phase in the same subspace, which persists during the whole measurement. We discussed how the latter can be determined from the full measurement time trace, using the method of state estimation. Quite surprisingly, we found that the outcome of this phase determination process is {\em independent} of the initial state in the state estimation procedure, making it useful in the situation of a parity {\em measurement}, where the initial state by definition is unknown. Finally, we show that the feed-back process is rather robust towards imperfections in the measurement record. This analysis opens up for realizing high-fidelity parity measurements in circuit QED, using realistic values of the coupling strength between the qubits and the cavity.
\section{Acknowledgments}\label{sec:ack}
We thank K. Lalumi\`ere, T. Stace and G. Milburn for valuable discussions. This work is supported by the European Commission through the
IST-015708 EuroSQIP integrated project and by the Swedish Research
Council.


\begin{thebibliography}{999}
\bibitem{stateref}
In the simulations the initial states were chosen as $\ket{\psi_+} = 1/\sqrt{2}(\ket{gg} + \ket{ee})$, $\ket{\psi_-} = 1/\sqrt{2}(\ket{ge} + \exp(i\pi/4)\ket{eg})$ and $\ket{\tilde{\psi}_-} = 1/\sqrt{2}(\ket{ge} + \ket{eg})$
\end{thebibliography}
\end{document}